\pdfoutput=1
\documentclass[%
preprint,
prb,
]{revtex4-1}

\usepackage{array}
\usepackage{calc}
\usepackage{etoolbox}
\robustify\bfseries
\usepackage{grffile}
\usepackage{graphicx}
\usepackage{color}
\usepackage{dcolumn}
\usepackage{bm}
\usepackage{multirow}
\usepackage{pbox}
\usepackage{tabularx,booktabs}
\usepackage{amsmath}
\usepackage{subfigure}
\usepackage{physics}
\usepackage{float}
\usepackage{inputenc}

\definecolor{pink}{rgb}{0.858, 0.188, 0.478}
\definecolor{green}{rgb}{0, 0.4, 0}

\makeatletter
\def\NAT@def@citea{\def\@citea{\NAT@separator}}
\makeatother

\begin{document}
\title{Band tail formation in mono and multilayered transition metal dichalcogenides: A detailed assessment and a quick-reference guide}

\author{Prasad Sarangapani}
\affiliation{School of Electrical and Computer Engineering, Purdue University, West Lafayette, IN 47907, USA}
\author{James Charles}
\affiliation{School of Electrical and Computer Engineering,
Purdue University, West Lafayette, IN 47907, USA}
\author{Tillmann Kubis}
\affiliation{School of Electrical and Computer Engineering, Purdue University, West Lafayette, IN 47907, USA}
\affiliation{Network for Computational Nanotechnology, Purdue University, West Lafayette, IN 47907, USA}
\affiliation{Purdue Center for Predictive Materials and Devices, Purdue University,  West Lafayette, IN 47907, USA}
\affiliation{Purdue Institute of Inflammation, Immunology and Infectious Disease, Purdue University,  West Lafayette, IN 47907, USA}
\affiliation{Purdue Quantum Science and Engineering Institute, West Lafayette, IN 47907, USA}

\email{prasad25081990@gmail.com}
\begin{abstract} 
Transition metal dichalcogenides (TMDs) are promising candidates for a wide variety of ultrascaled electronic, quantum computation, and optoelectronic applications. The exponential decay of electronic density of states into the bandgap, i.e. the band tail has a strong impact on the performance of TMD applications. In this work, the band tails of various TMD monolayer and multilayer systems when placed on various dielectric substrates is predicted with density functional theory based nonequilibrium Green's functions. Nonlocal scattering of electrons on polar optical phonons, charged impurities and remote scattering on phonons in the dielectric materials is included in the self-consistent Born approximation. The band tails are found to critically depend on the layer thickness, temperature, doping concentration and particularly on the chosen dielectric substrate. The underlying physical mechanisms are studied in high detail and an analytical interpolation formula is given to provide a quick-reference for Urbach parameters in MoS\textsubscript{2}, WS\textsubscript{2} and WSe\textsubscript{2}. 
\end{abstract}
\maketitle
\section{Introduction}
    Two dimensional materials have attracted considerable attention due to their unique electronic, optical and mechanical properties~\cite{novoselov2005two,wang2012electronics,jariwala2014emerging, mak2016photonics, manzeli20172d}. Transition metal dichalcogenides (TMDs) have a finite band gap making it an attractive alternative in electronics for Si/SiGe based transistors~\cite{desai2016mos2, podzorov2004high, liu2011performance, huyghebaert20182d}, in optoelectronics as possible materials in light emitting diodes~\cite{cheng2014electroluminescence, baugher2014optoelectronic, zhou20182d, gu2019room, bie2017mote, zheng2018light}  and solar cell applications~\cite{tsai2014monolayer, dasgupta2017thin, tsai2014monolayer, yang2014engineering}. TMD layers are coupled by weak van der Waals forces only which allows for mechanical cleavage of bulk TMD materials into mono and multilayer systems. Those systems yield electronic and optical properties that depend strongly on the number of layers~\cite{wang2017control}. Stacking multiple TMD layers on top of each other significantly widens the available material design space~\cite{pospischil2016optoelectronic, memaran2015pronounced} resulting in a plethora of ultra thin devices such as stacked optoelectronic p-n junctions~\cite{lee2014atomically, zhang2016van, mak2016photonics}, photovoltaics~\cite{yu2017layered, wu2019highly, das2019role} as well as ultra scaled non-volatile and neuromorphic memory devices~\cite{park2020nonvolatile, cao20202d, wang2019artificial}. Electrons in TMDs scatter on phonons, defects and charged impurities which leads to band tails (also known as Urbach tails), i.e. exponentially decaying density of states in the band gap~\cite{halperin1966impurity, sarangapani2019band}~\cite{schenk1998finite, sernelius1986band, halperin1966impurity, halperin1967impurity, van1992theory}. The slope of the exponential density of states tail is known as the Urbach parameter. Urbach tails can significantly alter the device performance: The switching of transistors is drastically affected by these tails~\cite{lu2014tunnel, agarwal2014band, bizindavyi2018band}. They affect the optical behaviour such as absorption spectra and absorption/recombination coefficients in optoelectronic devices~\cite{hebig2016optoelectronic, ikhmayies2013study, guerra2016urbach}. They also set a fundamental limit on the subthreshold performance of semiconductor devices at cryogenic temperatures for large-scale quantum computing applications.~\cite{beckers2019theoretical, beckers2021generalized} Since electron-phonon and electron-defect interaction cause the Urbach tails to form, the Urbach parameter is strongly dependent on temperature and doping concentration.~\cite{halperin1966impurity, halperin1967impurity, john1986theory,jain1991simple}. State-of-the-art models for the Urbach parameter of specific materials are either heuristic or the parameters are directly extracted from experimental observations.~\cite{halperin1966impurity, van1992theory, halperin1967impurity, jain1991simple, zhang2016effect, bizindavyi2018band, schenk1998finite, sernelius1986band}. TMD based nanodevices typically interface the TMD layers with various oxides. Therefore, TMD device electrons scatter on remote phonons as well~\cite{van2020effects, jena2007enhancement}. Experiments for several TMDs have shown that their Urbach tails depend on the oxide type~\cite{amani2015near,fang2014strong}. Different scattering mechanisms can interfere with each other and impact the electronic density of states which is hard to estimate without detailed calculations.~\cite{jirauschek2014modeling, eliashberg1960interactions, sarangapani2019band}. This is further complicated by the strong dependence of electronic and scattering properties in TMD systems on the number of atomic layers and their environment~\cite{kumar2012tunable,jena2007enhancement, ma2014charge}. Therefore, this work predicts Urbach parameters for MoS\textsubscript{2}, WS\textsubscript{2} and WSe\textsubscript{2} layer systems as a function of layer thickness, temperature, doping concentration, and oxide type (comprising Al\textsubscript{2}O\textsubscript{3}, HfO\textsubscript{2}, SiO\textsubscript{2} and h-BN) with DFT-based quantum transport calculations~\cite{szabo2015ab, afzalian2021ab} for electronic Green's functions including scattering on various types of phonons, charged impurities and remote scattering on oxide phonons with NEMO5.~\cite{charles2016incoherent, sarangapani2019band, lemus2020mode}. We delineate the contribution of different scattering mechanisms towards the Urbach parameter and gain insight into its dependence on layer thickness and oxide type. All calculations in this work are based on the non-equilibrium Green's function (NEGF) implementation of NEMO5. NEGF is well suited to analyze Urbach parameters~\cite{sarangapani2019band, khayer2011effects} since it is a method of choice to predict electronic behavior when incoherent scattering and coherent quantum effects are equally important. Electronic, thermal and optoelectronic systems with nanoscale dimensions or pronounced nonequilibrium conditions are a few such examples~\cite{lake1997single, markussen2009electron, lee2002nonequilibrium, kubis2009theory, sarangapani2018atomistic, yuanchen2019, geng2018quantitative, wang2020introduction}.This work summarizes the quantum transport calculations with an easily accessible lookup formula to predict Urbach parameters of MoS\textsubscript{2}, WS\textsubscript{2} and WSe\textsubscript{2} layer systems as a function of layer thickness, temperature, doping concentration, and oxide type.
\section{Simulation Approach}
\begin{figure}[H]
    \centering
    \includegraphics[scale=0.25]{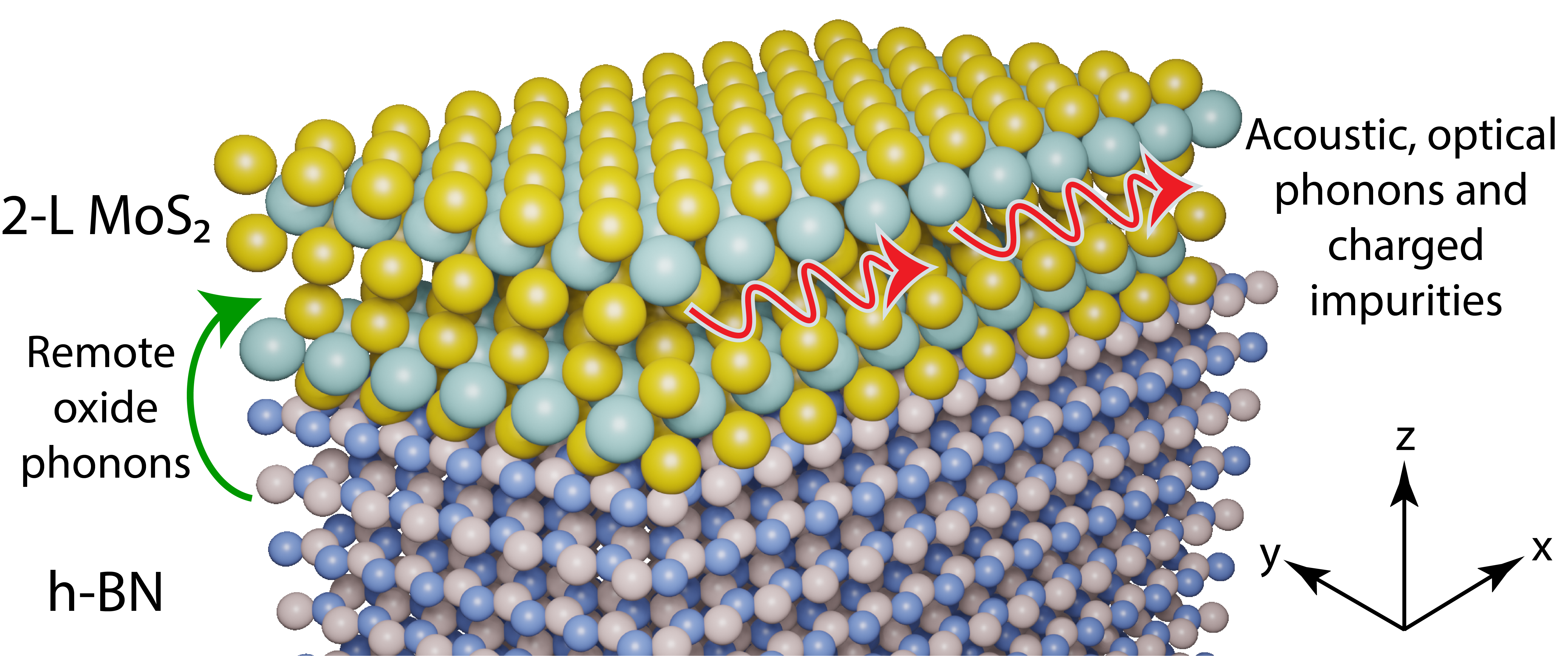}
    \caption{This representative schematic of the simulated TMD layers placed on top of oxides shows the atomic representation of bilayer MoS2 on top of hBN. Electrons are solved with scattering on acoustic and optical phonons and charged impurities within the TMD layers. In addition, electrons scattering remotely on oxide phonons is included. The x-direction and z-direction are resolved in real space, the y-direction is assumed to be periodic.}
    \label{fig:AtomicStructure}
    \end{figure}

     All TMD structures considered in this work are represented in their native atomic lattice. Electrons are sub-atomically resolved with maximally localized Wannier functions (MLWFs) derived from DFT Hamiltonians.~\cite{wang2017control, szabo2015ab}. The process of generating electronic Hamiltonian operators requires first to perform a self-consistent electronic structure calculation with the DFT tool VASP~\cite{hafner2008ab} with a convergence criterion of $10^{-8}$ eV. A momentum mesh of $5\times5\times5$ Monkhorst-Pack grid and an energy cutoff of 520 eV is used along with van-der-Waals force included following Ref.~\onlinecite{bucko2010improved}. The applied DFT model is the generalized gradient approximation (GGA) employing the Perdew-Burke-Ernzerhof (PBE) functionals. The resulting DFT Hamiltonian is then transformed into an MLWF representation using the Wannier90 software~\cite{marzari1997maximally, thygesen2005partly, das2017layer} with d orbitals for the metal sites and sp3 hybridized orbitals for the chalcogenide sites as the initial projection. The spreading of the Wannier functions is reduced iteratively until it converges to around $2A^2$. The atom positions and their corresponding electronic Hamiltonian of finite TMD structures are then imported into NEMO5 for NEGF calculations. All electronic Green's functions are solved in the self-consistent Born approximation with self-energies representing each incoherent scattering mechanism (scattering of electrons with acoustic phonons (AP), polar optical phonons (POP), charged impurities (CI) and remote phonons (RP) from the oxide. Since the structures are assumed to be periodic in the transverse Y direction (see Fig.~\ref{fig:AtomicStructure}), the electronic Green's functions and self-energies depend on the electronic energy $E$ and and in-plane momentum $k$. The corresponding Brillouin zone is resolved with 25 points. All Green's functions and self-energies are matrices with respect to the MLWF orbitals along the $\vec{x}= (X,Z)$ directions. (see Fig.~\ref{fig:AtomicStructure}). Each scattering processes is modeled with a corresponding retarded and lesser scattering self-energy~\cite{lake1997single, jirauschek2014modeling}. The imaginary part of the retarded self-energy provides information about the scattering rate of the electrons~\cite{wacker2002semiconductor}. The real part of the retarded self-energies yields an energy shift of electronic states~\cite{sarangapani2019band, esposito2009quantum}. Since this work focuses on the Urbach parameter only, the real part of all retarded scattering self-energies is ignored.
    
    
    In this work, the Green's functions of electrons are explicitly solved, whereas Green's functions of phonons are approximated as plane waves occupied with the equilibrium Bose distribution $N_{ph}$. The 3 acoustic phonon types (LA, TA, ZA) are averaged into a single effective deformation potential and sound velocity. The corresponding self-energy given in Ref.~\onlinecite{kubis2009theory} is multiplied by 3x accordingly. Polar optical phonons (POP) are modeled with a constant, material dependent phonon energy $\hbar\omega_{LO}$. 
    Scattering on charged impurities (CI) is assumed to be elastic. POP and CI are based on long-range Coulomb interaction and therefore yield non-local scattering self-energies.~\cite{mahan2013many, jacoboni2010theory} To limit the numerical burden of solving self-energies and Green's functions but still faithfully predict the POP and CI scattering, the respective self-energies are approximated to be local and multiplied with a material and device dependent compensation factor. The detailed expressions for POP and CI scattering self-energies and the compensation factor are given in Ref~\onlinecite{sarangapani2019band}. The interaction potential of electrons and remote oxide phonons is taken from Ref.~\onlinecite{fischetti2001effective}. The resulting lesser and retarded scattering self-energies for electrons remotely scattering on the two surface optical oxide phonon modes ($\nu = 1,2$) are given by 

    \begin{equation}
    \begin{array}{c} \label{eq:remotelessselfenergy2D}
        {\displaystyle \Sigma^{<}}\left(\vec{x}_{1},\vec{x}_{2},k,E\right)=\sum_{\nu}\dfrac{e^{2}}{\left(2\pi\right)}\dfrac{\hbar\omega_{SO,\nu}}{2\epsilon_{o}}{\displaystyle \left(\dfrac{1}{\epsilon_{ox}^{\infty}+\epsilon_{s}^{\infty}}-\dfrac{1}{\epsilon_{ox}^{s}+\epsilon_{s}^{\infty}}\right)}{\displaystyle \int dk'}I(k,k',\vec{x}_{1},\vec{x}_{2},z_{1},z_{2})\\
    \times\left[N_{ph}G^{<}\left(\vec{x}_{1},\vec{x}_{2},k',E-\hbar\omega_{SO,\nu}\right)+\left(1+N_{ph}\right)G^{<}\left(\vec{x}_{1},\vec{x}_{2},k',E+\hbar\omega_{SO,\nu}\right)\right]
    \end{array}
    \end{equation}

    \begin{equation} \label{eq:remoteretselfenergy2D}
    \begin{array}{c}
    {\displaystyle \Sigma^{R}\left(\vec{x}_{1},\vec{x}_{2},k,E\right)}=\sum_{\nu}\dfrac{e^{2}}{\left(2\pi\right)}\dfrac{\hbar\omega_{SO,\nu}}{2\epsilon_{o}}{\displaystyle \left(\dfrac{1}{\epsilon_{ox}^{\infty}+\epsilon_{s}^{\infty}}-\dfrac{1}{\epsilon_{ox}^{s}+\epsilon_{s}^{\infty}}\right)}{\displaystyle \int dk'}I(k,k',\vec{x}_{1},\vec{x}_{2},z_{1},z_{2})\\
    \times\left[(1+N_{ph})G^{R}\left(\vec{x}_{1},\vec{x}_{2},l,E-\hbar\omega_{SO,\nu}\right)+N_{ph}G^{R}\left(\vec{x}_{1},\vec{x}_{2},k',E+\hbar\omega_{SO,\nu}\right)+\dfrac{1}{2}G^{<}(\vec{x}_{1},\vec{x}_{2},k',E-\hbar\omega_{SO,\nu})\right.\\
    \left.-\dfrac{1}{2}G^{<}(\vec{x}_{1},\vec{x}_{2},k',E+\hbar\omega_{SO,\nu})+i{\displaystyle \int}\dfrac{d\tilde{E}}{2\pi}G^{<}\left(\vec{x}_{1},\vec{x}_{2},k',\tilde{E}\right)\left(Pr\dfrac{1}{E-\tilde{E}-\hbar\omega_{SO,\nu}}-Pr\dfrac{1}{E-\tilde{E}+\hbar\omega_{SO,\nu}}\right)\right]
    \end{array}
    \end{equation}

    where 

    \begin{equation}
    I(k,k',\vec{x}_{1},\vec{x}_{2},z_{1},z_{2})=
    2\pi J_{0}\left(\sqrt{\left(k - k'\right)^2 + \zeta^{-2}}\left|\vec{x}_{1}-\vec{x}_{2}\right|\right)e^{-\sqrt{\left(k - k'\right)^2 + \zeta^{-2}}\left(z_{1}+z_{2}-2t\right)}
    \end{equation}
    $\hbar\omega_{SO,\nu}$ is the $\nu^{th}$ optical phonon frequency of the underlying oxide, $\epsilon_{ox}^{s}$ and $\epsilon_{ox}^{\infty}$  are the static and infinite frequency dielectric constants of the oxide, $\epsilon_s^{\infty}$ is the infinite frequency dielectric constant of the TMD. The oxide phonon modes are considered to exponentially decay into the TMD with $z_1$ and $z_2$ the distances of the two electron propagation coordinates from the oxide-semiconductor interface (see Fig.~\ref{fig:AtomicStructure}). $t$ is the thickness of the TMD system. $J_0$ is the Bessel-J function of 0'th order. 
    The electrostatic screening of electron and holes is represented by $\zeta$ and calculated with the Lindhard formula~\cite{lindhard1954properties} (Add the 2D formula with more details)
    \begin{equation}
    \begin{array}{c}
    \zeta_{Lindhard}=\left(\dfrac{e^{2}}{\epsilon_{o}\epsilon_{s}}\dfrac{-2}{(2\pi)^{3}}{\displaystyle \int}d\vec{q}\left.\dfrac{\partial f}{\partial\epsilon}\right|_{\epsilon(\vec{q})}\right)^{-1/2}
    \end{array}
    \label{eq:Lindhard}
    \end{equation}
    where $f$ is the Fermi distribution function and the momentum integral runs over the first Brillouin zone.

   For all the discussions in the subsequent sections, the electrons are solved in equilibrium. The electronic Fermi level is determined such that the spatially integrated electron density agrees with the integrated doping concentration to achieve global charge neutrality. The Green's functions are solved with the Dyson and Keldysh equations: 
    
    \begin{equation}
    \begin{array}{c}
    G^{R}=\left(EI-H-\Sigma_{AP}^R - \Sigma_{POP}^{R}-\Sigma_{CI}^{R}-\Sigma_{ROP}^{R}-\Sigma_{source}^{R}-\Sigma_{drain}^{R}\right)^{-1}\\
    G^{<}=G^{R}\left(\Sigma_{AP}^{<} + \Sigma_{POP}^{<}+\Sigma_{CI}^{<}+\Sigma_{ROP}^{<}+\Sigma_{source}^{<}+\Sigma_{drain}^{<}\right)G^{R\dagger}\\
    \end{array}
    \end{equation}

All scattering self-energies are self-consistently solved with the Green's functions until the relative particle current variation is less than 1e-5 throughout the device. The source and drain contact self-energies are solved following Ref.~\cite{sancho1985highly}. Urbach parameter is extracted from the exponentially decaying spatially averaged density of states below (above) the conduction (valence) band.~\cite{sarangapani2019band} 
Those Urbach parameters correspond to the ones measured in transport experiments. In contrast, optical measurements relate to excitons which require a different scattering model. 
    All other material parameters for the TMDs and oxides are taken from Refs~\cite{kumar2012tunable, jin2014intrinsic, ma2014charge} and listed in Table~\ref{table:TMDMaterialParams} and \ref{table:OxideParams}.

    \begin{table}
    \begin{centering}
   \begin{tabular}{|c|c|c|c|c|c|c|c|c|}
    \hline 
    Material & Layer & $v_{s}$(m/s) & $\rho$(kg/m\textsuperscript{3}) & $\hbar\omega_{LO}$(meV) & D (eV/nm) & $\epsilon_{s}$  & $\epsilon_{\infty}$ & POP scattering prefactor\tabularnewline
    \hline 
    \multirow{4}{*}{$\mathrm{MoS_{2}}$ } & 1 & \multirow{4}{*}{7200} & \multirow{4}{*}{5060} & \multirow{4}{*}{48} & 4.5 & 3.8  & 3.2  & 0.4069\tabularnewline
    \cline{2-2} \cline{6-9} \cline{7-9} \cline{8-9} \cline{9-9} 
     & 2 &  &  &  & 5.37 & 5.65  & 4.8  & 0.2584\tabularnewline
    \cline{2-2} \cline{6-9} \cline{7-9} \cline{8-9} \cline{9-9} 
     & 3 &  &  &  & 6.25 & 6.47  & 5.5  & 0.2248\tabularnewline
    \cline{2-2} \cline{6-9} \cline{7-9} \cline{8-9} \cline{9-9} 
     & 4  &  &  &  & 7.12 & 7.3  & 6.2  & 0.2004\tabularnewline
    \hline 
    \multirow{4}{*}{$\mathrm{WS_{2}}$ } & 1 & \multirow{4}{*}{6670} & \multirow{4}{*}{7500} & \multirow{4}{*}{33} & 3.2 & 3.65  & 3.1  & 0.5847\tabularnewline
    \cline{2-2} \cline{6-9} \cline{7-9} \cline{8-9} \cline{9-9} 
     & 2  &  &  &  & 4.07 & 5.15  & 4.37  & 0.4169\tabularnewline
    \cline{2-2} \cline{6-9} \cline{7-9} \cline{8-9} \cline{9-9} 
     & 3  &  &  &  & 4.95 & 5.92  & 5.03  & 0.3595\tabularnewline
    \cline{2-2} \cline{6-9} \cline{7-9} \cline{8-9} \cline{9-9} 
     & 4 &  &  &  & 5.82 & 6.7  & 5.69  & 0.3187\tabularnewline
    \hline 
    \multirow{4}{*}{$\mathrm{WSe_{2}}$ } & 1 & \multirow{4}{*}{5550} & \multirow{4}{*}{9320} & \multirow{4}{*}{30} & 3.2 & 3.7  & 3.145  & 0.6167\tabularnewline
    \cline{2-2} \cline{6-9} \cline{7-9} \cline{8-9} \cline{9-9} 
     & 2 &  &  &  & 4.07 & 5.3  & 4.5  & 0.4337\tabularnewline
    \cline{2-2} \cline{6-9} \cline{7-9} \cline{8-9} \cline{9-9} 
     & 3 &  &  &  & 4.95 & 6.1  & 5.18  & 0.3765\tabularnewline
    \cline{2-2} \cline{6-9} \cline{7-9} \cline{8-9} \cline{9-9} 
     & 4 &  &  &  & 5.82 & 6.9  & 5.86  & 0.3326\tabularnewline
    \hline 
    \end{tabular}\protect\caption{Sound velocity $v_s$, material density $\rho$, LO phonon frequency ($\hbar\omega_{LO}$), deformation potentials (D) and dielectric constants ($\epsilon_s$ and $\epsilon_\infty$) of for 1-4 layers of MoS\textsubscript{2}, WS\textsubscript{2} and WSe\textsubscript{2} used in this work. Parameters have been taken from Refs~\cite{kumar2012tunable} and \cite{jin2014intrinsic}.}
    \label{table:TMDMaterialParams}
    \par\end{centering}
    \end{table}

    \begin{table}
    \begin{centering}
    \begin{tabular}{|c|c|c|c|c|c|}
    \hline 
    Oxide & $\hbar\omega_{SO,1}$ (meV) & $\hbar\omega_{SO,2}$ (meV) & $\epsilon_{s}$ & $\epsilon_{\infty}$ & ROP scattering prefactor\tabularnewline
    \hline 
    $h-BN$ & 93.07 & 179.10 & 5.09 & 4.10 & 0.1159\tabularnewline
    \hline 
    $SiO_{2}$ & 55.60 & 138.10 & 3.90 & 2.50 & 1.0478\tabularnewline
    \hline 
    $Al_{2}O_{3}$ & 48.18 & 71.41 & 12.53 & 3.20 & 2.9231\tabularnewline
    \hline 
    $HfO_{2}$ & 12.40 & 48.35 & 23.00 & 5.03 & 3.6363\tabularnewline
    \hline 
    \end{tabular}\protect\caption{Soft optical phonon frequencies ($\hbar\omega_{SO, 1}$ and $\hbar\omega_{SO, 2}$), static and infinite frequency dielectric constants of the oxides used in this work taken from Ref~\cite{ma2014charge}.}
    \label{table:OxideParams}
    \par\end{centering}
    \end{table}

\section{Results}
\subsection{Band tails: Intrinsic to the material}
   
    \begin{figure}[H]
    \centering
    \includegraphics[scale=0.8]{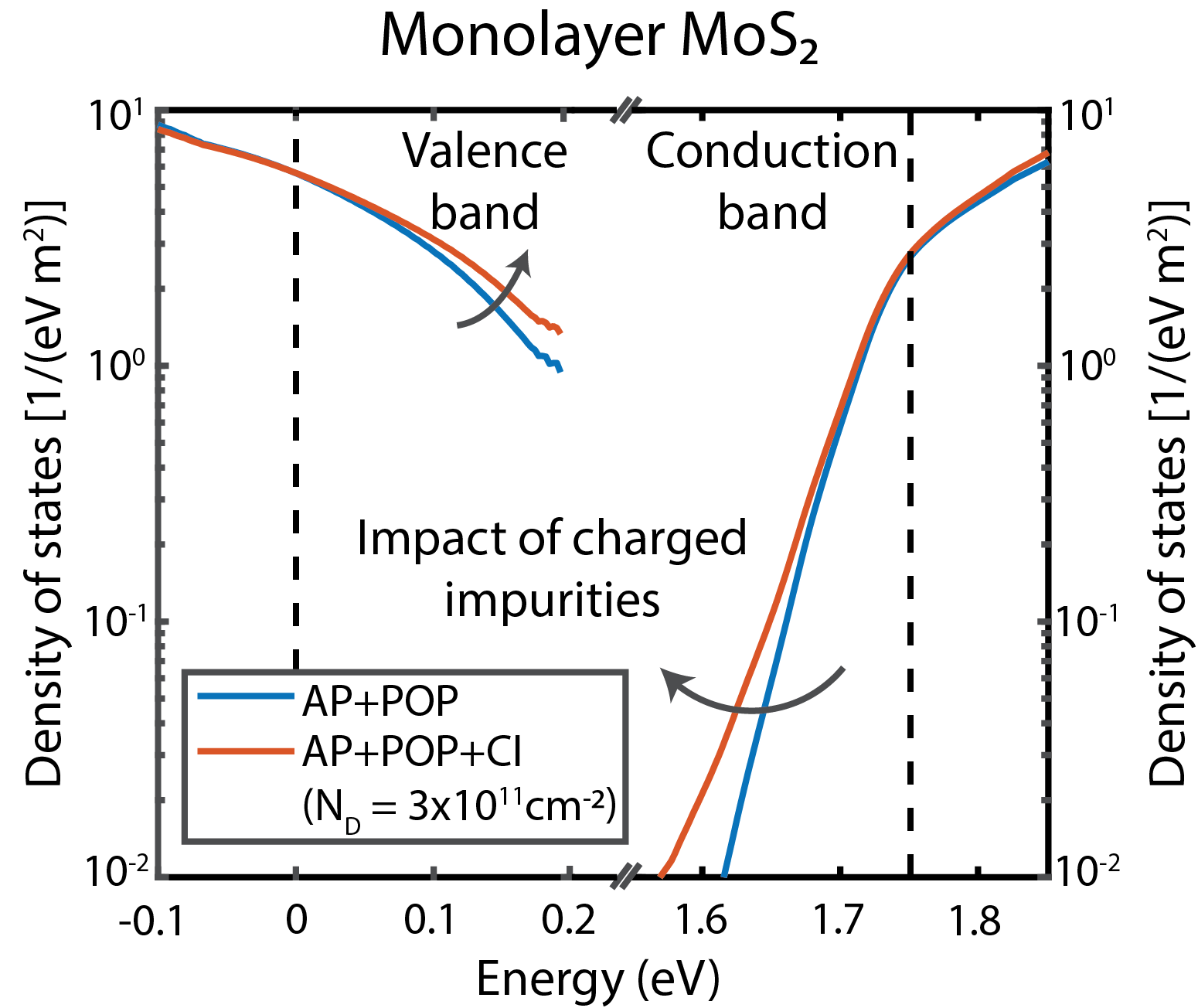}
    \caption{Density of states of monolayer MoS\textsubscript{2} close to valence and conduction band edges. Scattering on acoustic phonons (AP) and polar optical phonons (POP) creates Urbach tails into the band gap (blue). Scattering on charged impurities (CI) (red) increases the Urbach tail further. The black dashed lines indicate the valence and conduction band edges without incoherent scattering.}
    \label{fig:MoS2DoSVBCBPOPImp}
    \end{figure}

\begin{figure}[H]
    \centering
    \includegraphics[scale=0.5]{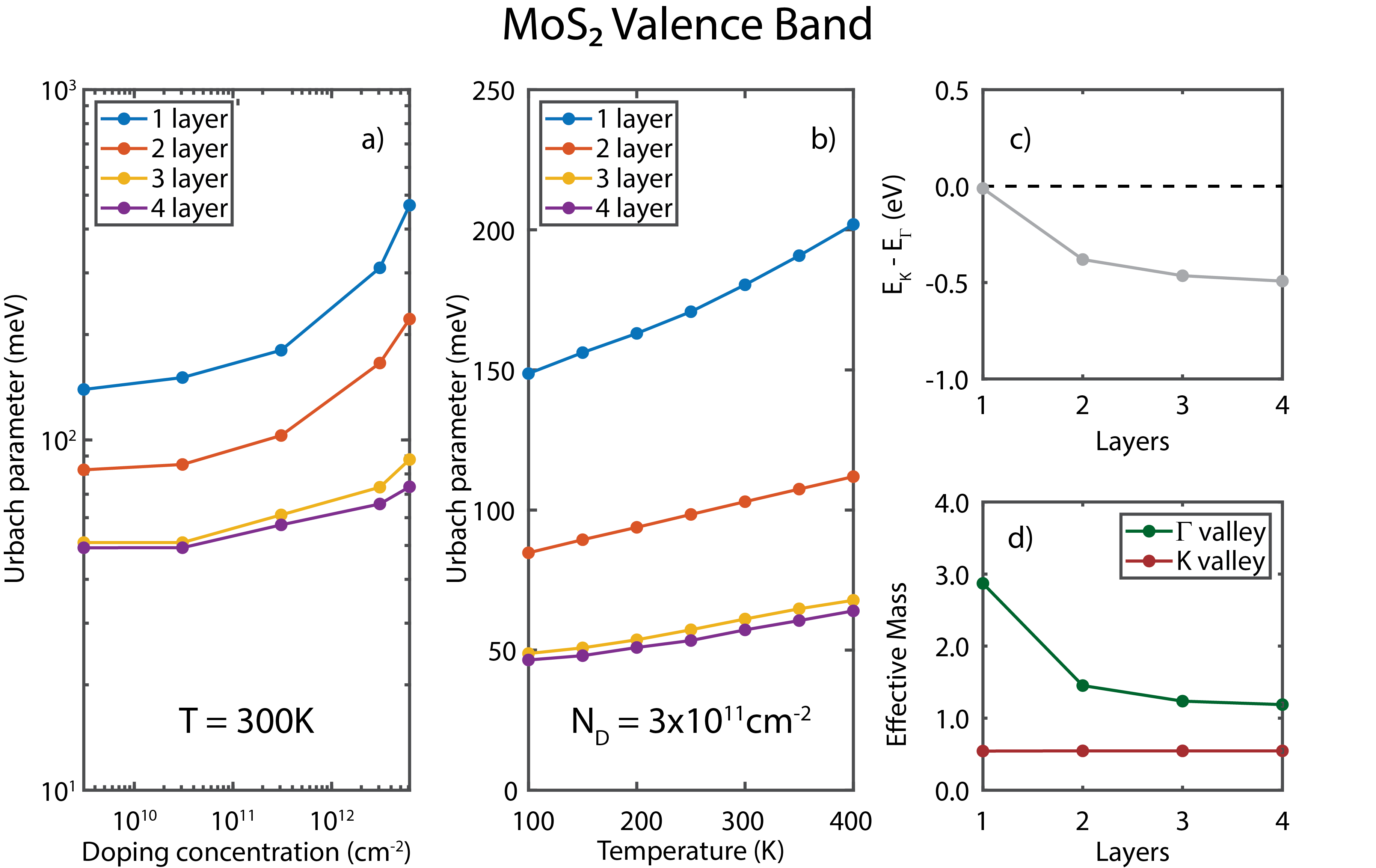}
    \caption{a) Urbach parameter of the MoS\textsubscript{2} valence band as a function of doping concentration for different number of layers at 300K. With increasing doping concentration, the Urbach parameter increases due to stronger impurity scattering. b) Urbach parameter of the MoS\textsubscript{2} valence band as a function of temperature for different number of layers at a doping concentration of $3\times10^{11}cm^{-2}$. With increasing temperature, the Urbach parameter increases due to stronger phonon scattering. Both a) and b) show smaller Urbach parameters with larger number of layers because of the reduction of valence band edge density of states with more layers: Only monolayer MoS\textsubscript{2} exhibits degenerate valence band K and $\Gamma$ valleys as depicted with their energy differences in c). The density of states of the $\Gamma$ valley is proportional to its effective mass which reduces with the number of MoS\textsubscript{2} layers, shown in d).}
    \label{fig:MoS2UrbachValenceBand}
    \end{figure}

    Electronic scattering on polar optical phonons (POP) creates finite density of states above (below) the valence (the conduction) band that decays exponentially into the band gap (see Fig.~\ref{fig:MoS2DoSVBCBPOPImp}). The slope of this band tail, i.e. the Urbach parameter increases  with incoherent scattering of electrons on charged impurities (as also observed in Ref.~\onlinecite{sarangapani2019band}). 
    The scattering rates are proportional to the imaginary retarded scattering self-energies ($\Sigma^R$)\cite{wacker2002semiconductor}. 
    In the self-consistent Born approximation $\Sigma^R$ is proportional to the retarded Green's function ($G^R$) (see Eq.~1 and Refs.~\cite{sarangapani2019band, charles2016incoherent}.). 
    Therefore, the strength of the scattering processes that form Urbach tails is determined by the imaginary part of $G^R$, i.e. the density of states at the band edges~\cite{sarangapani2019band}.
    This is the root cause for the effective mass dependency of scattering rates in Fermi Golden rule models (see e.g. \cite{jacoboni2010theory, mahan2013many}).
    Since the valence band of monolayer MoS\textsubscript{2} has a much larger effective mass ($2.886*m_e$) than its conduction band ($0.596*m_e$)~\cite{wang2017control} the scattering strength, and with it the Urbach parameter is larger for the holes (see Fig.~\ref{fig:MoS2DoSVBCBPOPImp}).
    The phonon (charged impurity) scattering self-energies are proportional to the phonon number (doping concentration)~\cite{sarangapani2019band}. 
    Accordingly, Figs.~\ref{fig:MoS2UrbachValenceBand} a) and b) show the Urbach parameters of MoS\textsubscript{2} valence band electrons increase with doping concentration and temperature. 
    
    Figures~\ref{fig:MoS2UrbachValenceBand} a) and b) also show a reduction of the Urbach parameter with the number of MoS\textsubscript{2} layers. This is due to the fact the density of states at the top of the valence band of monolayer MoS\textsubscript{2} is larger than the ones of any multilayer MoS\textsubscript{2} system. Particularly the degeneracy of K and $\Gamma$ valleys is lifted as soon as more than one layer of MoS\textsubscript{2} is present (see Fig.~\ref{fig:MoS2UrbachValenceBand} c)). That is why adding a second MoS\textsubscript{2} layer gives the largest reduction in the Urbach parameter.
    The effective mass of the $\Gamma$ valley, i.e. the highest valence band valley of multilayer MoS\textsubscript{2} systems, and with it the density of states at the top of the valence band declines continuously with the number of layers (see Fig.~\ref{fig:MoS2UrbachValenceBand} d)).
    In addition, the polar optical phonon scattering potential decreases with larger dielectric constants~\cite{sarangapani2019band, kubis2009theory, ridley2013quantum} which were observed to increase in thicker MoS\textsubscript{2} layers~\cite{kumar2012tunable, jin2014intrinsic}. 
    It is worth to mention, we have also observed decreasing impact of scattering in thicker ultrathin bodies of III-V materials in the Ref.~\onlinecite{sarangapani2019band}.
    
    \begin{figure}[H]
    \centering
    \includegraphics[scale=0.5]{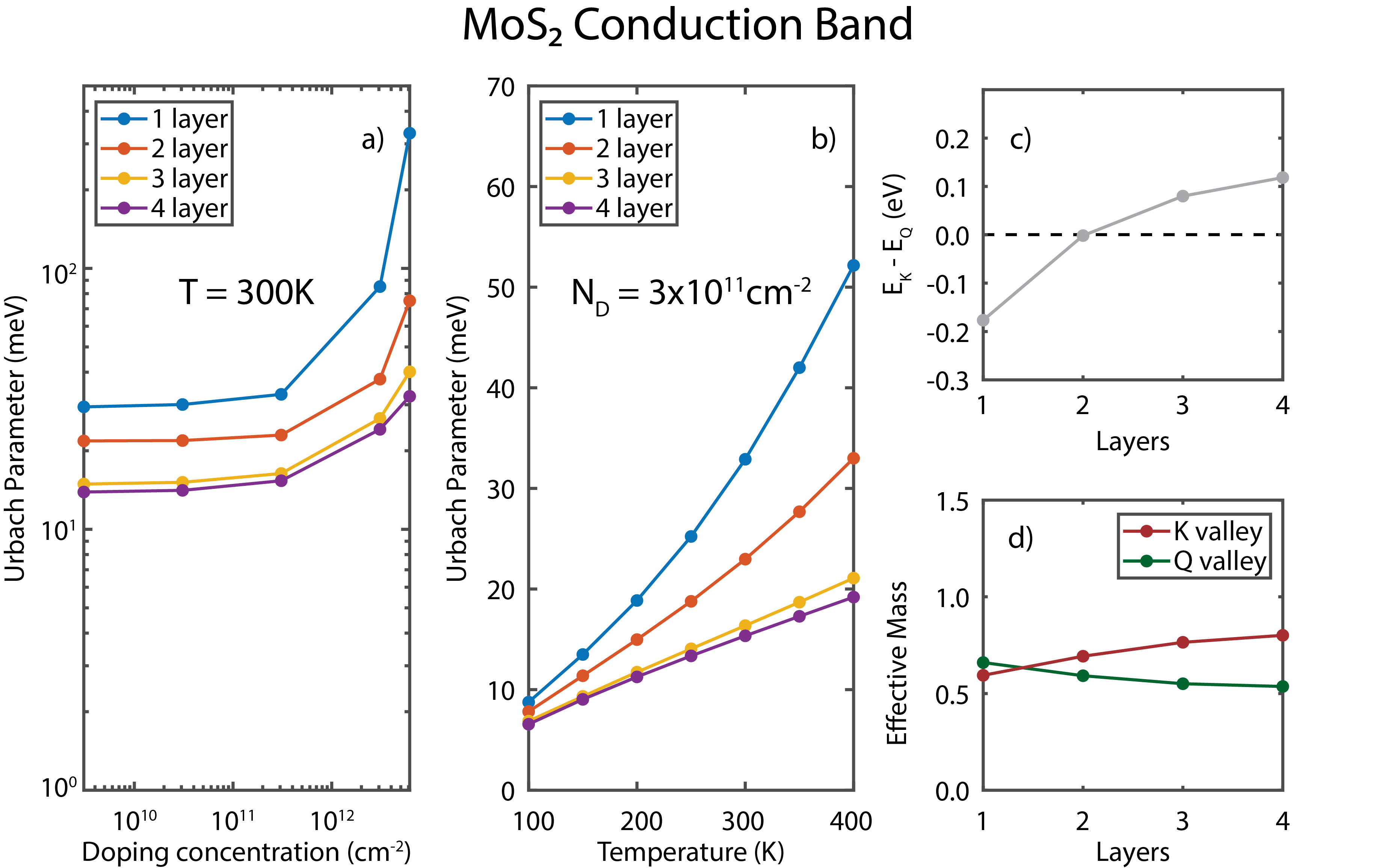}
    \caption{Similar as in Figs.~\ref{fig:MoS2UrbachValenceBand}, the Urbach parameter of the MoS\textsubscript{2} conduction band is a function of doping concentration a) and temperature b). Differing from the valence band results of Figs.~\ref{fig:MoS2UrbachValenceBand}, the K and Q valleys are degenerate in the bilayer case, see c) and the conduction band edge effective mass has a relatively small layer thickness dependence shown in d). In spite of the bilayer valley degeneracy, the Urbach parameter still decreases monotonously with the number of layers, since the reduction in scattering potential of the bilayer overcompensates its increase in band edge density of states (see table~\ref{table:TMDMaterialParams}).}
    \label{fig:MoS2UrbachConductionBand}
    \end{figure}

   The conduction band of MoS\textsubscript{2} has a similar dependence on doping concentration and temperature as the valence band (see Figs.~\ref{fig:MoS2UrbachConductionBand} a) and b)). Different to the valence band, however, the conduction band valleys K and Q are degenerate only in the bilayer case (see Fig.~\ref{fig:MoS2UrbachConductionBand} c)). In spite of the expected increase in band edge density of states and with it an increase of the conduction band Urbach parameter of the bilayer MoS\textsubscript{2}, the Urbach parameter shows a monotonous decrease with the number of MoS\textsubscript{2} layers (see Figs.~\ref{fig:MoS2UrbachConductionBand} a) and b)). This is due to a significant reduction of the calculated scattering self-energy prefactor of polar optical phonon scattering from monolayer to bilayer MoS\textsubscript{2} (see table~\ref{table:TMDMaterialParams}). 
    Overall, the conduction band Urbach parameters of MoS\textsubscript{2} layers are lower than those of the valence band due the lower conduction band effective masses (see Fig.~\ref{fig:MoS2UrbachConductionBand} d)).

    \begin{figure}[H]
    \centering
    \includegraphics[scale=0.5]{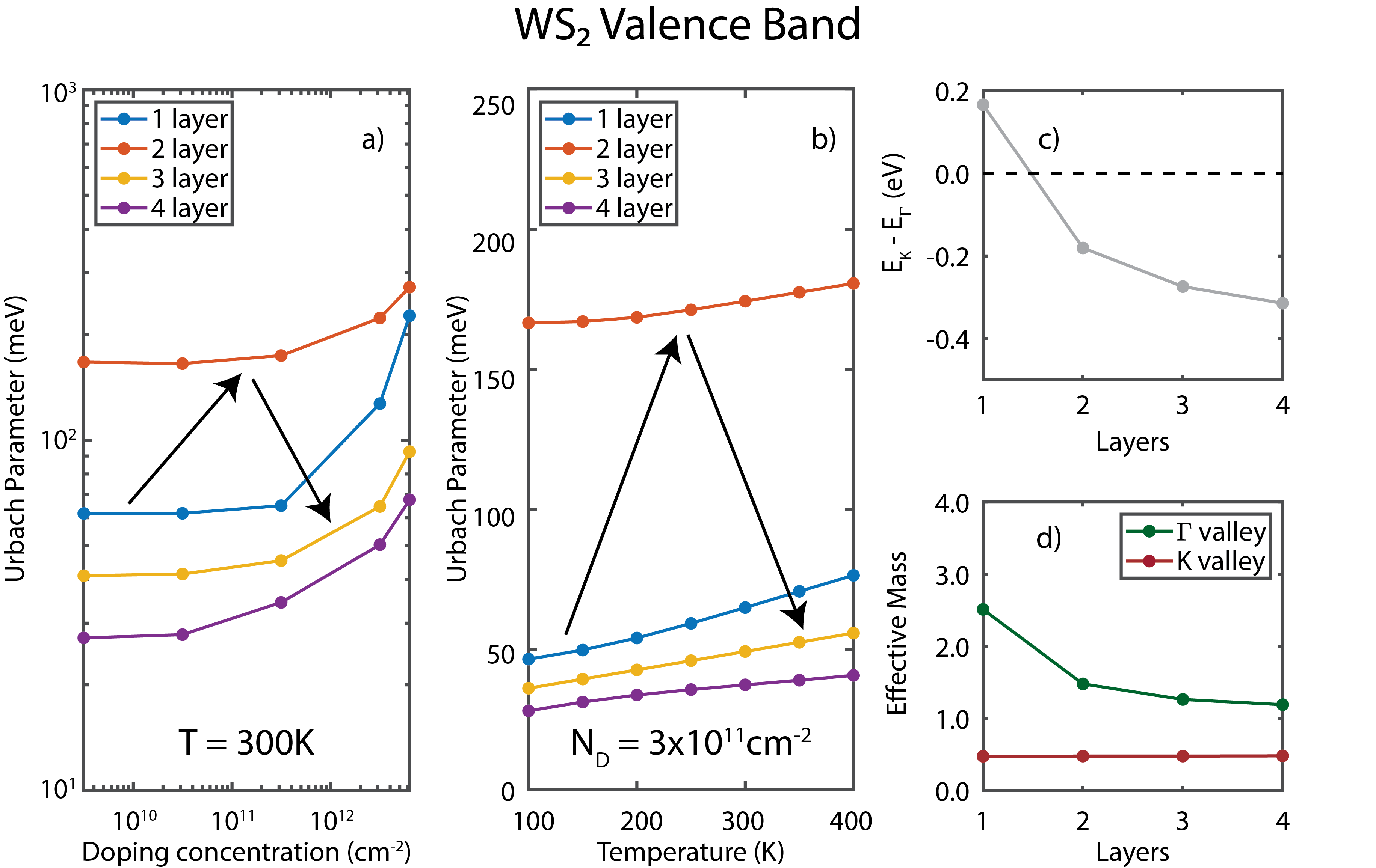}
    \caption{a) Urbach parameter of the WS\textsubscript{2} valence band as a function of doping concentration for different number of layers at 300K (a) and as a function of temperature for a doping density of $3\times10^{11}cm^{-2}$. Similar to the valence band Urbach parameter of MoS\textsubscript{2} (Figs.~\ref{fig:MoS2UrbachValenceBand}), the Urbach parameter of WS\textsubscript{2} increases with doping concentration and temperature. The energy of K-valleys exceed those of the $\Gamma$ valleys only for monolayer WS\textsubscript{2} as shown in c). The transition of the valence band edge from K to $Gamma$ valley and the large difference of K and $\Gamma$ valley effective masses (shown in d)) cause a maximum of the WS\textsubscript{2} band edge density of states and accordingly a maximum of the Urbach parameter for the bilayer configuration (indicated with arrows).}
    \label{fig:WS2UrbachValenceBand}
    \end{figure}
    
    The valence band Urbach parameters of WS\textsubscript{2} layers in Figs.~\ref{fig:WS2UrbachValenceBand} a) and b) show similar changes with doping concentration and temperature as the MoS\textsubscript{2} valence band results of Figs.~\ref{fig:MoS2UrbachValenceBand}). However, the Urbach parameter is largest for bilayer WS\textsubscript{2}. This is due to a transition of the valence band edge from K to $\Gamma$ valley when more than one layer of WS\textsubscript{2} is present (see Fig.~\ref{fig:WS2UrbachValenceBand} c)). Given the large difference of K and $\Gamma$ valley effective masses (see Fig.~\ref{fig:WS2UrbachValenceBand} d)), this transition causes a maximum in the valence band edge density of states and therefore (as discussed above) a maximum in the scattering strength for bilayer WS\textsubscript{2}. Accordingly the Urbach parameter follows this trend (indicated with arrows in Figs.~\ref{fig:WS2UrbachValenceBand} a) and b). WS\textsubscript{2} with more than 2 layers show the similar reduction in the Urbach parameter as discussed already for MoS\textsubscript{2} in Figs.~\ref{fig:MoS2UrbachValenceBand}.
    
    \begin{figure}[H]
    \centering
    \includegraphics[scale=0.5]{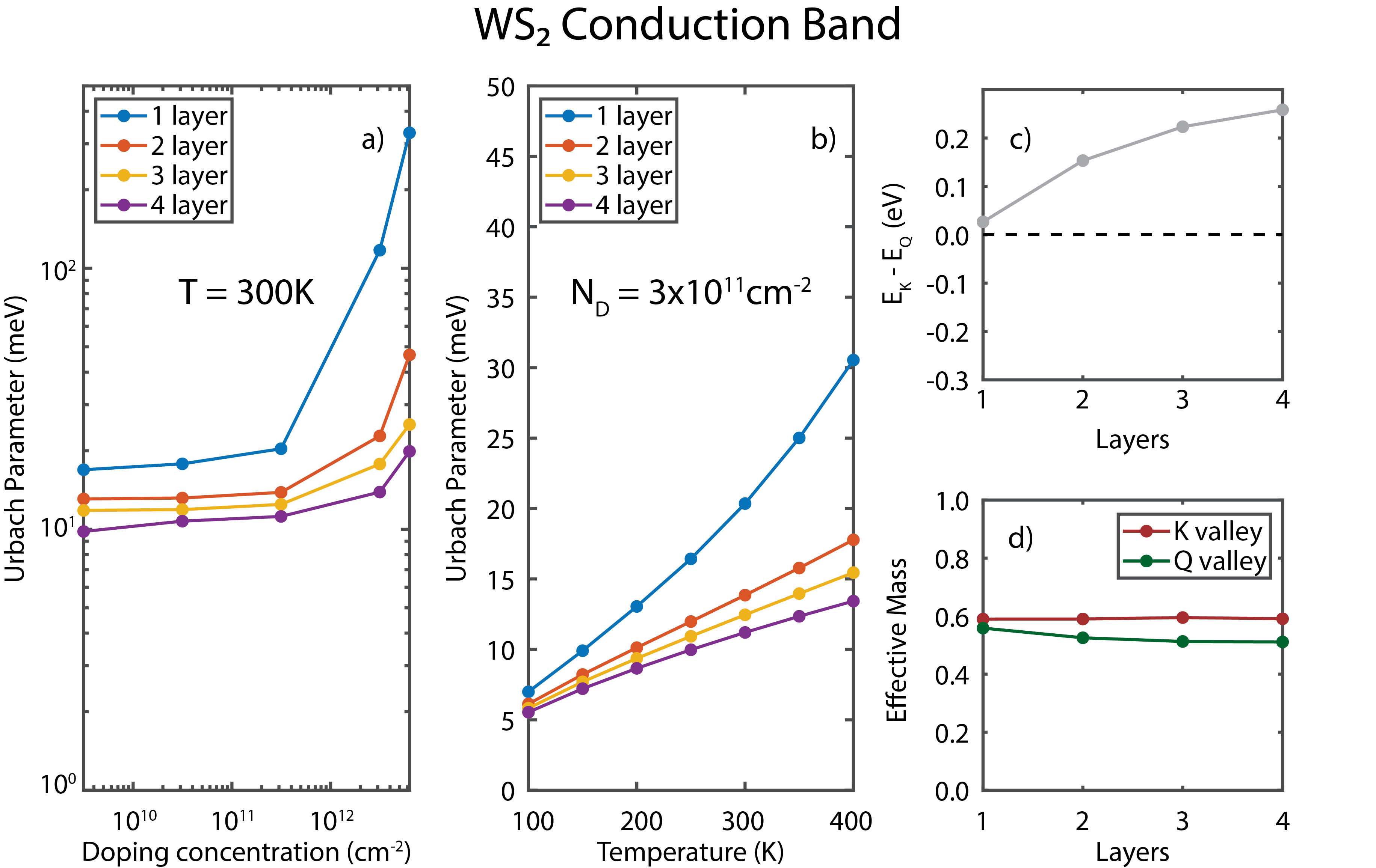}
    \caption{a) 
    The Urbach parameter of the WS\textsubscript{2} conduction band is a function of doping concentration a) and temperature b) similar for the conduction band of MoS\textsubscript{2} in Figs.~\ref{fig:MoS2UrbachConductionBand}. The Urbach parameter declines with increasing number of layers mainly due to the reduction of the scattering potential (see table \ref{table:TMDMaterialParams}), since neither valley degeneracy in c) nor the valley effective mass in d) changes significantly with the number of layers.}
    \label{fig:WS2UrbachConductionBand}
    \end{figure}
    
\begin{figure}[H]
\centering
\includegraphics[scale=0.5]{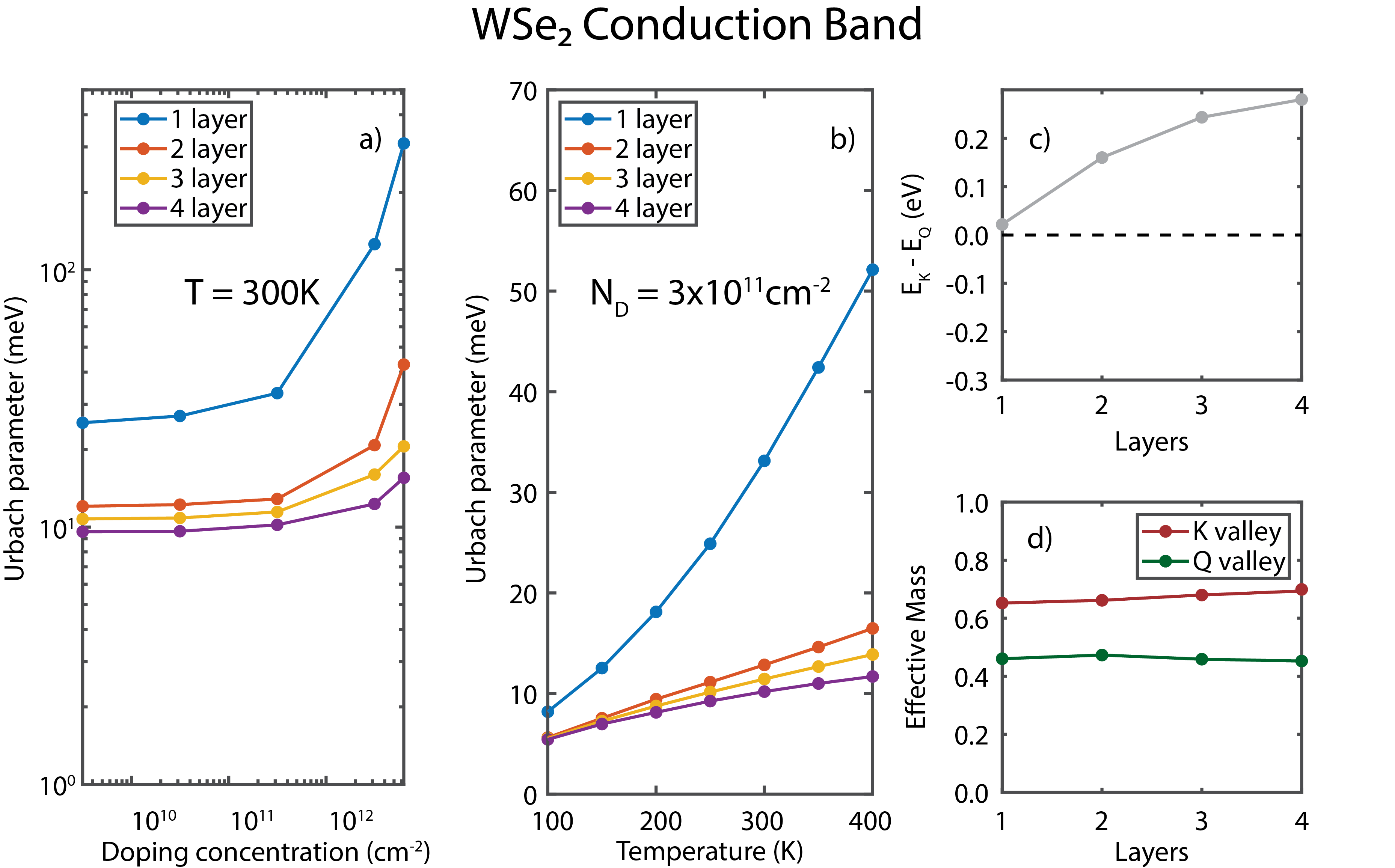}
\caption{a) The Urbach parameter of the WSe\textsubscript{2} conduction band is a function of doping concentration a) and temperature b) similar to the conduction band of WSe\textsubscript{2} in Figs.~\ref{fig:WSe2UrbachConductionBand}. The Urbach parameter declines with increasing number of layers mainly due to the reduction of the scattering potential (see table \ref{table:TMDMaterialParams}), since neither valley degeneracy in c) nor the valley effective mass in d) changes significantly with the number of layers.}
\label{fig:WSe2UrbachConductionBand}
\end{figure}

   The conduction band Urbach parameter of WS\textsubscript{2} (see Figs.~\ref{fig:WS2UrbachConductionBand}) is very similar to the one of MoS\textsubscript{2} (see Figs.~\ref{fig:MoS2UrbachConductionBand}) in its overall dependence on doping and temperature. The scattering potential of conduction band of WS\textsubscript{2} and with it the Urbach parameter decays with the number of layers (see table~\ref{table:TMDMaterialParams}). Neither significant effective mass changes nor valley degeneracies influence this behavior for WS\textsubscript{2} conduction bands. The same is true for valence and conduction bands of WSe\textsubscript{2} layers (see Supplementary Figs.~\ref{fig:WSe2UrbachValenceBand} and ~\ref{fig:WSe2UrbachConductionBand}).  
    
    \begin{figure}[H]
    \centering
    \includegraphics[scale=0.8]{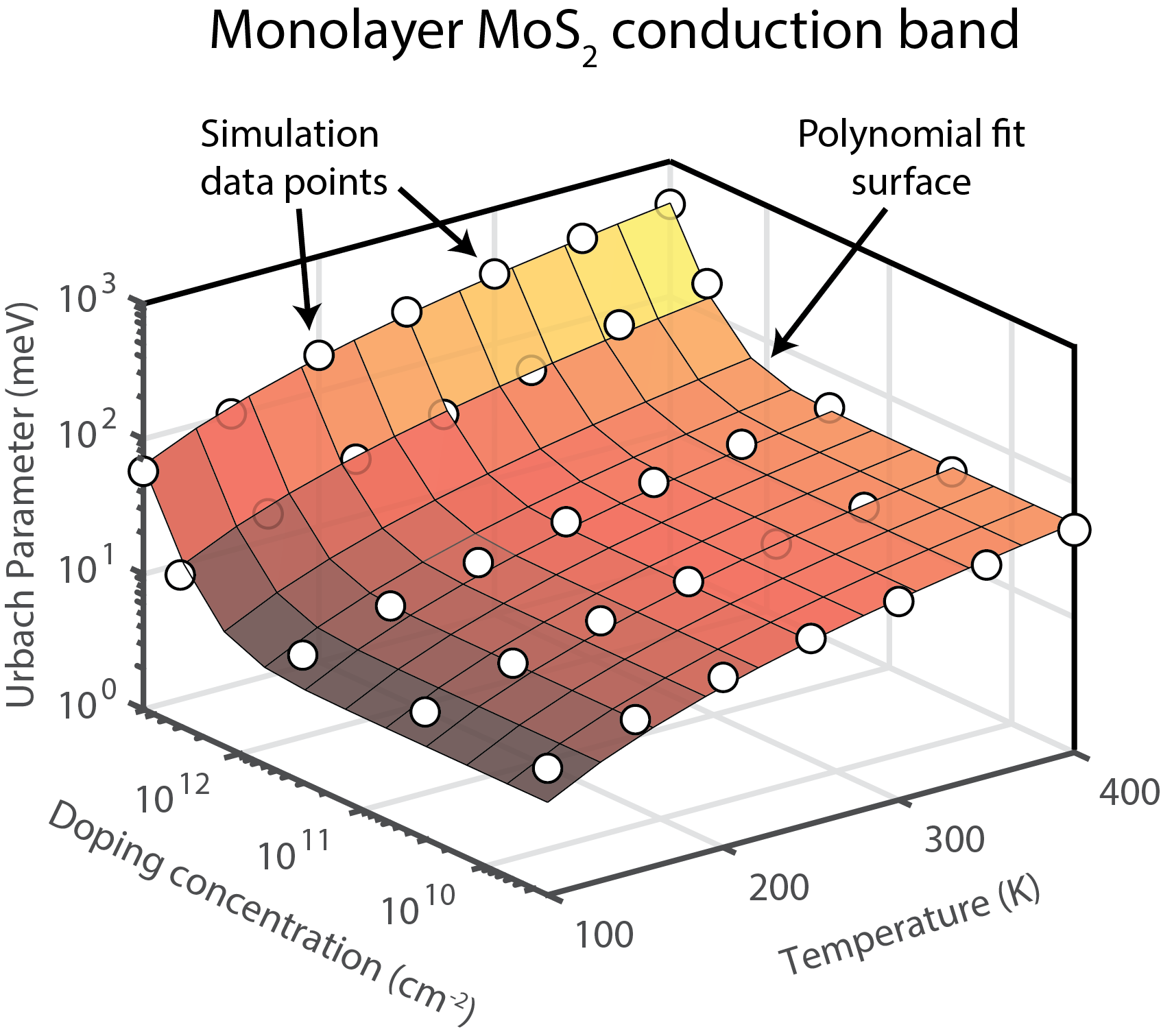}
    \caption{Urbach parameter as a function of doping concentration and temperature for monolayer MoS\textsubscript{2}. The white circles represent the simulation data points and the contour represents its least square polynomial fit. The fit parameters for this and all other systems are given in table~\ref{table:FitParams}}
    \label{fig:MoS2UrbachContourPlot}
    \end{figure}
    
   For completeness and to ease estimating the Urbach parameter as a function of doping concentration, temperature, layer thickness and material, all calculated Urbach parameters have been input to a polynomial fit given by 
    \begin{equation}
        Urbach(T, N_D) = a(T/300)^p + b(T/300)^q(N_D/1e12)^r
    \label{eq:fittingEquation}
    \end{equation}
    Figure~\ref{fig:MoS2UrbachContourPlot} exemplifies this fit for the conduction band Urbach parameter of MoS\textsubscript{2}.
    Since the Urbach parameter depends non-monotonically on the number of layers, the least-square fitting is performed for each layer separately. 
    The fit parameters of all consider TMD systems together with their $R^2$ fit values can be found in table~\ref{table:FitParams}.
    
    \subsection{Band tails: Impacted by dielectric materials}
\begin{figure}[H]
    \centering
    \includegraphics[scale=0.8]{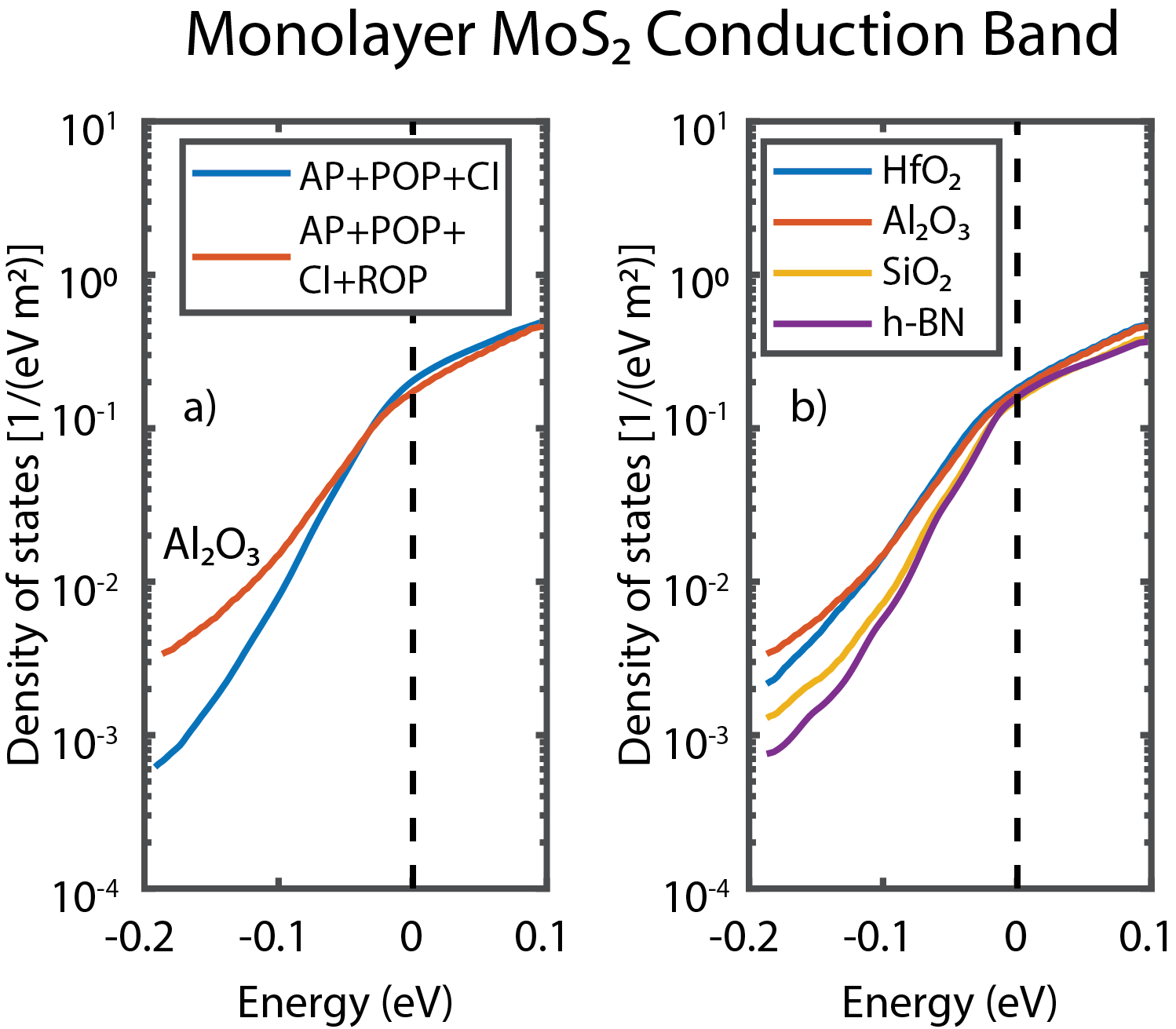}
    \caption{Density of states of monolayer MoS\textsubscript{2} when exfoliated on various dielectric materials, close to valence and conduction band edges (solid lines). The remote scattering on the dielectric material phonons increases the Urbach tail compared to the intrinsic value of monolayer MoS\textsubscript{2} (dotted lines). The impact of the dielectric material phonon scattering follows the same order that is given in table~\ref{table:OxideParams}, i.e. the strength of the remote phonon scattering prefactor. The black dashed lines indicate the valence and conduction band edges without incoherent scattering.}
    

    \label{fig:MoS2DOSRemote}
    \end{figure}

    In typical TMD based nanodevices, the TMD layers are capped with dielectric materials. The remote scattering on phonons in those dielectric materials contribute significantly to the Urbach band tail, as shown in Fig.~\ref{fig:MoS2DOSRemote}. 
    
    The relative impact of remote phonon scattering on phonons in the dielectric materials is directly proportional to the scattering self-energy prefactor listed in table~\ref{table:OxideParams}. This prefactor is determined by the difference of the dielectric constants and the energies of the soft optical phonon modes (see Eq.~\ref{eq:remoteretselfenergy2D}). It is worth to mention, HfO\textsubscript{2} has a large scattering impact given its high difference in dielectric constants and the comparably low soft optical phonon mode energies. Following the same arguments, the low scattering contributions of hBN and SiO2 originate in their high phonon mode energies and small difference in their dielectric constants.
    
    \begin{figure}[H]
    \centering
    \includegraphics[scale=0.6]{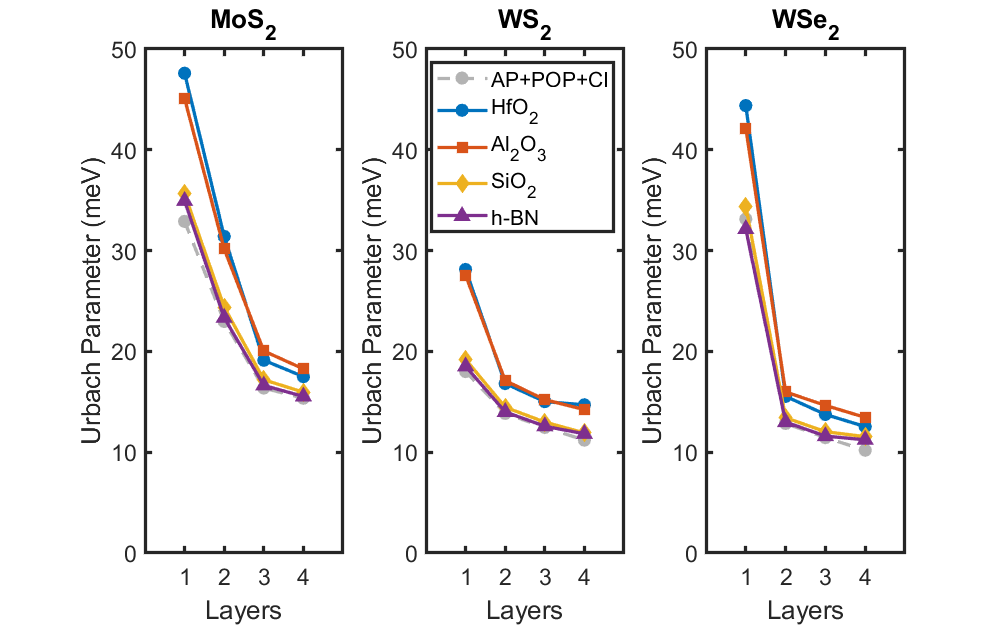}
    \caption{Urbach parameter for a) MoS\textsubscript{2}, b) WS\textsubscript{2} and c) WSe\textsubscript{2} systems capped with 4 different dielectric insulators - HfO\textsubscript{2} (dots), Al\textsubscript{2}O\textsubscript{3} (squares), SiO\textsubscript{2} (diamonds) and h-BN (triangles) along with their intrinsic values shown in open symbols. With larger TMD layer thickness, the maximum of the electron density is located farther from the semiconductor-insulator interface. Therefore, the Urbach parameter gets less affected by the insulator for thicker TMD systems. As seen in Fig.~\ref{fig:MoS2DOSRemote}, HfO\textsubscript{2} and Al\textsubscript{2}O\textsubscript{3} increase the Urbach parameter most due to their large scattering potentials.}
    \label{fig:UrbachVsLayersOxides}
    \end{figure}
    
 
   Figure~\ref{fig:UrbachVsLayersOxides} shows the RP scattering impacts the Urbach parameter more in thinner TMD systems than in the thicker ones. This is due to the smaller spatial confinement in larger TMD systems which allows the charge density in the TMD layers to move farther away from the insulator-TMD interface. Accordingly, the difference in the contributions of the various insulators decreases with increasing TMD thickness.
    Beyond 3 layers, the Urbach parameter starts to saturate and slowly approaches the intrinsic value of the respective TMD. As seen already in Fig.~\ref{fig:MoS2DOSRemote}, the Urbach parameters are consistently higher for Al\textsubscript{2}O\textsubscript{3} and HfO\textsubscript{2} due to their strong RP scattering potentials. 

\section{Conclusion}
Quantum transport calculations of electrons in TMD systems are used to predict the formation of band tails due to scattering on phonons, charged impurities and remote scattering on substrate dielectric phonons. 
All materials are atomically resolved, electronic Hamiltonian operators are based on DFT, and incoherent, non-local scattering is modeled in self-consistent Born approximation. 
It is shown that the Urbach band tail parameter strongly depends on temperature, impurity concentration and TMD layer thickness as well as on the type of dielectric substrate. 
The details of the Urbach parameter depend on a balance of valley degeneracies, scattering potentials and phonon occupancies. 
This can result in non-monotonic behavior of the Urbach parameter as seen e.g. in WS\textsubscript{2}. 
To ease reproducibility of the sophisticated quantum transport calculations, we present analytical approximations to the observed Urbach parameter predictions. 
Among the dielectric materials considered, Al\textsubscript{2}O\textsubscript{3} and HfO\textsubscript{2} are shown to contribute strongest to the remote phonon scattering enhanced band tails. 
With increasing TMD layer count, electrons spread farther away from the dielectric and thus the contribution of remote scattering on dielectric phonons decreases.

\section{Acknowledgements}
James Charles and Tillmann Kubis acknowledge support by the Silvaco Inc. This research
was supported in part through computational resources provided by Information Technology at Purdue, West
Lafayette, Indiana. The authors acknowledge the Texas Advanced Computing Center (TACC) at The University
of Texas at Austin for providing HPC resources that have contributed to the research results reported within this paper. This research used resources of the Oak Ridge Leadership Computing Facility, which is a DOE Office of Science User Facility supported under Contract DEAC05-00OR22725. Prasad Sarangapani would like to acknowledge Purnima Padmanabhan for helping with the figures.

\section{Supplementary Information}
\begin{table}[H]
\begin{centering}
\begin{tabular}{|c|c|c|c|c|c|c|c|c|}
\hline 
 & Band Type & Layers & $a$ & $b$ & $p$ & $q$ & $r$ & $R^{2}$ fit value\tabularnewline
\hline 
\multirow{8}{*}{$MoS_{2}$} & \multirow{4}{*}{Valence} & 1 & 149.57 & 61.162 & 0.1461 & 0.8073 & 0.9078 & 0.9955\tabularnewline
\cline{3-9} \cline{4-9} \cline{5-9} \cline{6-9} \cline{7-9} \cline{8-9} \cline{9-9} 
 &  & 2 & 82.981 & 39.460 & 0.1116 & 0.73851 & 0.6793 & 0.9978\tabularnewline
\cline{3-9} \cline{4-9} \cline{5-9} \cline{6-9} \cline{7-9} \cline{8-9} \cline{9-9} 
 &  & 3 & 50.096 & 14.623 & 0.1014 & 1.3003 & 0.4714 & 0.9817\tabularnewline
\cline{3-9} \cline{4-9} \cline{5-9} \cline{6-9} \cline{7-9} \cline{8-9} \cline{9-9} 
 &  & 4 & 49.101 & 0.1101 & 0.1002 & 3.7729 & 2.3337 & 0.9271\tabularnewline
\cline{2-9} \cline{3-9} \cline{4-9} \cline{5-9} \cline{6-9} \cline{7-9} \cline{8-9} \cline{9-9} 
 & \multirow{4}{*}{Conduction} & 1 & 30.567 & 5.7612  & 1.6752  & 1.4527  & 2.1574 & 0.9965\tabularnewline
\cline{3-9} \cline{4-9} \cline{5-9} \cline{6-9} \cline{7-9} \cline{8-9} \cline{9-9} 
 &  & 2 & 22.461  & 2.4506 & 1.0747 & 1.6788 & 1.6752 & 0.9988\tabularnewline
\cline{3-9} \cline{4-9} \cline{5-9} \cline{6-9} \cline{7-9} \cline{8-9} \cline{9-9} 
 &  & 3 & 15.368  & 2.9972 & 0.8734 & 0.0885 & 1.1907 & 0.9923\tabularnewline
\cline{3-9} \cline{4-9} \cline{5-9} \cline{6-9} \cline{7-9} \cline{8-9} \cline{9-9} 
 &  & 4 & 48.497  & 1.8010 & 0.0590 & 6.5438 & 0.3661 & 0.9009\tabularnewline
\hline 
\multirow{8}{*}{$WS_{2}$} & \multirow{4}{*}{Valence} & 1 & 63.210 & 11.144 & 0.4238 & 1.3126 & 1.5879 & 0.9951\tabularnewline
\cline{3-9} \cline{4-9} \cline{5-9} \cline{6-9} \cline{7-9} \cline{8-9} \cline{9-9} 
 &  & 2 & 172.43  & 18.337 & 0.1102 & 0.4899 & 0.8087 & 0.9222\tabularnewline
\cline{3-9} \cline{4-9} \cline{5-9} \cline{6-9} \cline{7-9} \cline{8-9} \cline{9-9} 
 &  & 3 & 40.343  & 12.360 & 0.2296 & 0.9083 & 0.4475 & 0.9848\tabularnewline
\cline{3-9} \cline{4-9} \cline{5-9} \cline{6-9} \cline{7-9} \cline{8-9} \cline{9-9} 
 &  & 4 & 27.545  & 13.102 & 0.1408 & 0.8728 & 0.5770 & 0.9782\tabularnewline
\cline{2-9} \cline{3-9} \cline{4-9} \cline{5-9} \cline{6-9} \cline{7-9} \cline{8-9} \cline{9-9} 
 & \multirow{4}{*}{Conduction} & 1 & 17.814  & 15.216 & 1.4801 & 1.0604 & 1.6482 & 0.9982\tabularnewline
\cline{3-9} \cline{4-9} \cline{5-9} \cline{6-9} \cline{7-9} \cline{8-9} \cline{9-9} 
 &  & 2 & 13.449  & 1.2230 & 0.7822 & 0.8175 & 1.8035 & 0.9970\tabularnewline
\cline{3-9} \cline{4-9} \cline{5-9} \cline{6-9} \cline{7-9} \cline{8-9} \cline{9-9} 
 &  & 3 & 12.046 & 1.3597 & 0.7226 & 0.0675 & 1.2610 & 0.9930\tabularnewline
\cline{3-9} \cline{4-9} \cline{5-9} \cline{6-9} \cline{7-9} \cline{8-9} \cline{9-9} 
 &  & 4 & 10.196  & 1.0997 & 0.4620 & 0 & 1.2088 & 0.9721\tabularnewline
\hline 
\multirow{8}{*}{$WSe_{2}$} & \multirow{4}{*}{Valence} & 1 & 45.762  & 4.3223 & 0.3396 & 1.0847 & 1.4928 & 0.9797\tabularnewline
\cline{3-9} \cline{4-9} \cline{5-9} \cline{6-9} \cline{7-9} \cline{8-9} \cline{9-9} 
 &  & 2 & 33.271 & 0.8838  & 0.2733 & 0 & 1.1190 & 0.9129\tabularnewline
\cline{3-9} \cline{4-9} \cline{5-9} \cline{6-9} \cline{7-9} \cline{8-9} \cline{9-9} 
 &  & 3 & 28.843 & 0.0825  & 0.2060 & 3.0934 & 2.8990 & 0.9810\tabularnewline
\cline{3-9} \cline{4-9} \cline{5-9} \cline{6-9} \cline{7-9} \cline{8-9} \cline{9-9} 
 &  & 4 & 24.699 & 0.5682 & 0.2214 & 1.7593 & 1.6473 & 0.9564\tabularnewline
\cline{2-9} \cline{3-9} \cline{4-9} \cline{5-9} \cline{6-9} \cline{7-9} \cline{8-9} \cline{9-9} 
 & \multirow{4}{*}{Conduction} & 1 & 28.203  & 18.465 & 1.4141 & 1.0200 & 1.4832 & 0.9990\tabularnewline
\cline{3-9} \cline{4-9} \cline{5-9} \cline{6-9} \cline{7-9} \cline{8-9} \cline{9-9} 
 &  & 2 & 12.450  & 1.0889 & 0.7411 & 0.8994 & 1.8314 & 0.9936\tabularnewline
\cline{3-9} \cline{4-9} \cline{5-9} \cline{6-9} \cline{7-9} \cline{8-9} \cline{9-9} 
 &  & 3 & 10.684  & 2.1195 & 0.6217 & 0 & 0.8655 & 0.9917\tabularnewline
\cline{3-9} \cline{4-9} \cline{5-9} \cline{6-9} \cline{7-9} \cline{8-9} \cline{9-9} 
 &  & 4 & 9.2162 & 1.6687 & 0.3946 & 0  & 0.7384 & 0.9311\tabularnewline
\hline 
\end{tabular}
\label{table:FitParams}
\par\end{centering}
\protect\caption{Parameters for predicting the Urbach tail with Eq.~\ref{eq:fittingEquation} as a function of temperature and doping concentration for MoS\textsubscript{2}, WS\textsubscript{2} and WSe\textsubscript{2} along with the $R^2$ deviation from the calculated quantum transport result. The parameters were determined with MATLAB Curve Fitting Toolbox~\cite{matlabcurvefit}.}
\end{table}

\begin{figure}[H]
\centering
\includegraphics[scale=0.5]{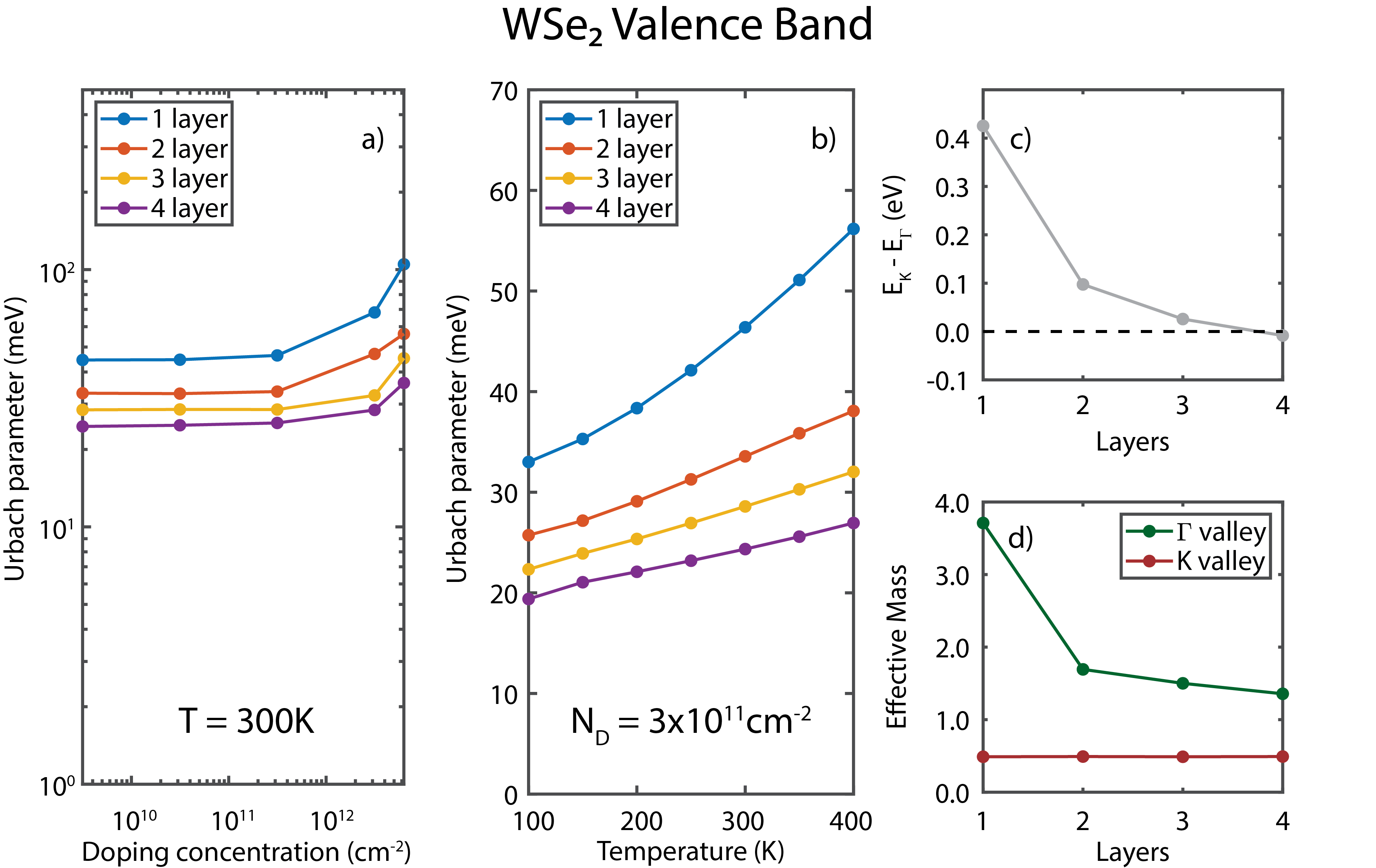}
\caption{a) Urbach parameter of the WSe\textsubscript{2} valence band as a function of doping concentration for different number of layers at 300K. With increasing doping concentration, the Urbach parameter increases due to stronger impurity scattering. b) Urbach parameter of the WSe\textsubscript{2} valence band as a function of temperature for different number of layers at a doping concentration of $3\times10^{11}cm^{-2}$. With increasing temperature, the Urbach parameter increases due to stronger phonon scattering. Both a) and b) show smaller Urbach parameters with larger number of layers because of the reduction of valence band edge density of states with more layers: Only 4-layer WSe\textsubscript{2} exhibits degenerate valence band K and $\Gamma$ valleys as depicted with their energy differences in c). Otherwise, the $\Gamma$ valley only contributes to the conduction. The density of states of the $\Gamma$ valley is proportional to its effective mass which reduces with the number of WSe\textsubscript{2} layers, shown in d).}
\label{fig:WSe2UrbachValenceBand}
\end{figure}


\bibliography{refs}

\end{document}